\documentclass[11pt]{panacm}

\usepackage{graphicx}
\usepackage{amsmath}
\usepackage{amsfonts}
\usepackage{amssymb}

\title{Droplet in micro-channels: a numerical approach using an adaptive two phase flow solver
}

\author{Jose-Maria Fullana, Yue Ling, St\'ephane Popinet, Christophe Josserand}

\heading{J-M Fullana, Y. Ling, S. Popinet, Ch. Josserand}

\address{$^{1}$ Sorbonne Universités, \\UPMC Univ Paris 06, UMR 7190, 
Institut Jean Le Rond d'Alembert, \\F-75005, Paris, France
\and
$^{2}$ CNRS, UMR 7190, \\ Institut Jean Le Rond d'Alembert, \\ F-75005, Paris, France	
4 Place Jussieu, Paris, France \\
jose.fullana@upmc.fr
}


\abstract{
We propose a numerical approach to study the mechanics of a flowing bubble in a constraint micro channel. Using an open source two phase flow solver (Gerris, gfs.sourceforge.net) we compute solutions of the bubble dynamics (i.e. shape and terminal velocity) induced by the interaction between the bubble movement,  the Laplace pressure variation, and the lubrication film near the channel wall. Quantitative and qualitative results are presented and compared against both theory and experimental data for small Capillary numbers. 
We discuss the technical issues of explicit integration methods on small Capillary numbers computations, and the possibility of adding Van der Walls forces to give a more precise picture of the Droplet-based microfluidic problem.
 }

\begin{document}
\maketitle

\section{INTRODUCTION}
Droplet-based microfluidics is a very promising tool for performing biochemical or chemical assays. Droplets are unit systems of controlled volume and content, within which mixing can be easily achieved. Several physical phenomena (mechanics, thermocapillarity, solutocapillarity, thermomechanics) either in cumulative or compensative ways appears when we develop microfluidic setups. It is of prime importance to characterize, under controlled experimental conditions, within which range each contribution is the dominant phenomena regarding element migration. Rationalizing these various effects would have important consequences for lab-on-a-chips, and numerical studies are an interesting way to understand each contribution separately.

In microfluidic setups we often have measurements of shape deformations, bubble or drop velocities, and pressures or flow rates at the input/exit conditions, but the knowledge local values of these variables are not easily available, because principally of the small length scales involved in the system. Numerical approaches are then an interesting way for acceding to small length dynamical fields. The validity of numerical approaches requires  the validation  and the confrontation against  theories and experimental data. Theoretical results exist in few academic microfluidic configurations like bubble in cylinders or plates, and it is then necessary to be able to compare positively these academic configurations before extend the prediction to more complex situations which are actually common in microfluidic devices.

In this communication we present an open source two phase flow tool (gfs.sourceforge.net) for computing solutions of the bubble dynamics (i.e. shape and terminal velocity)  in microfluidic channel and; numerical results are presented and compared, quantitatively  and qualitatively, against both theory and experimental data for small Capillary numbers. 
We discuss the possibility of adding Van der Walls forces to give a more precise picture of the Droplet-based microfluidic problem.

\section{EQUATIONS AND NUMERICAL SCHEME}

We use the incompressible, two-dimensional variable-density, Navier–Stokes equations with surface tension which can be written
\begin{eqnarray}
  \label{eq:5}
\rho (  {\partial U \over \partial t} + U \nabla U )= - \nabla p + \mu \nabla^2 U + \sigma \kappa \delta {\bar n} \\
\nabla U=0 \nonumber
\end{eqnarray}
with $U = (u, v)$ the fluid velocity, $\rho = \rho(x, t)$ the fluid density, $\mu = \mu(x, t)$ the dynamic viscosity. The Dirac distribution function $\delta$ expresses the fact that the surface tension term is concentrated on the interface; $\sigma$ is the surface tension coefficient, $\kappa$ and ${\bar n}$ the curvature and normal to the interface.

For two-phase flows we introduce the volume fraction $c(x,t)$ of the first fluid and define the density and viscosity as a function of $c$, i.e. $\rho = \rho(c(x,t))$ and $\mu=\mu(c(x,t))$. The advection equation for the density can then be replaced with an equivalent advection equation for the volume fraction
\begin{eqnarray*}
\partial_t c + \nabla (U c) =0
\end{eqnarray*}

The Navier-Stokes equations are solved using a finite volume approach based into a projection method. The numerical problem is solved using the open-source package Gerris \cite{OpenSource}. A staggered in time discretisation of the volume-fraction/density and pressure leads to the a formally second-order accurate time discretisation. The interface between the different fluids are tracked and followed using a VOF (Volume of Fluid) method. The spatial discretisation is done using a quad-tree square cells which give a very important flexibility allowing dynamical costless grid refinement into user-defined regions. Finally a powerful discretisation scheme was developed to capture accurately the surface tension term. More information can be found in reference \cite{Popinet}

\section{BUBBLE IN A CHANNEL}

\begin{figure}[htb]
\centering
\includegraphics[width=8cm]{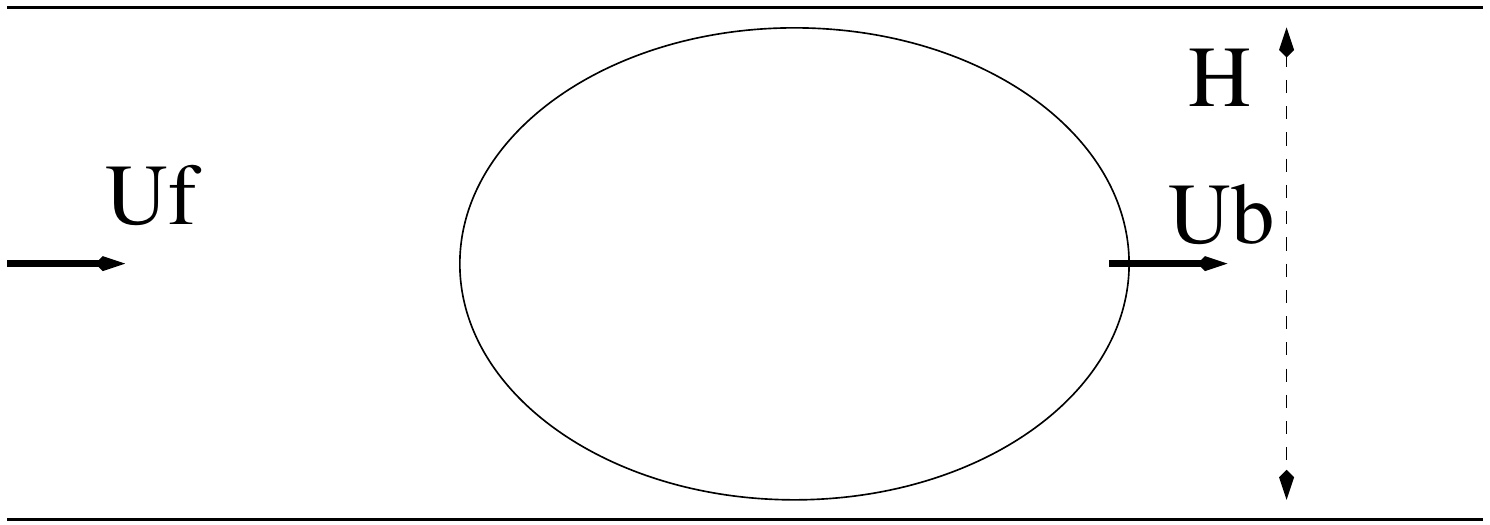}
\caption{Microfluid configuration : channel width is $H$, velocities are $U_f$ the mean flow velocity and $U_b$ the bubble velocity.}
\label{fig1}
\end{figure}

The studied configuration is presented in Figure \ref{fig1}: a bubble is pushed into a  microchannel of width $H$ by a mean flow velocity $U_f$. The typical length of the bubble is larger than the width $H$, the bubble is then constrained by the channel. In the stationary regime the velocity bubble is $U_b$. Using the width $H$ as characteristic length and the mean flow velocity $U_f$ as characteristic velocity the dimensionless Navier-Stokes is then written

\begin{figure}[htb]
\centering
\includegraphics[width=0.4\textwidth]{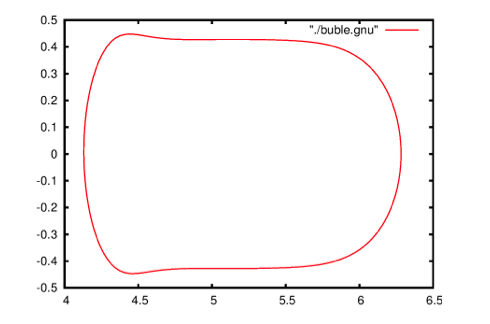}\includegraphics[width=0.5\textwidth]{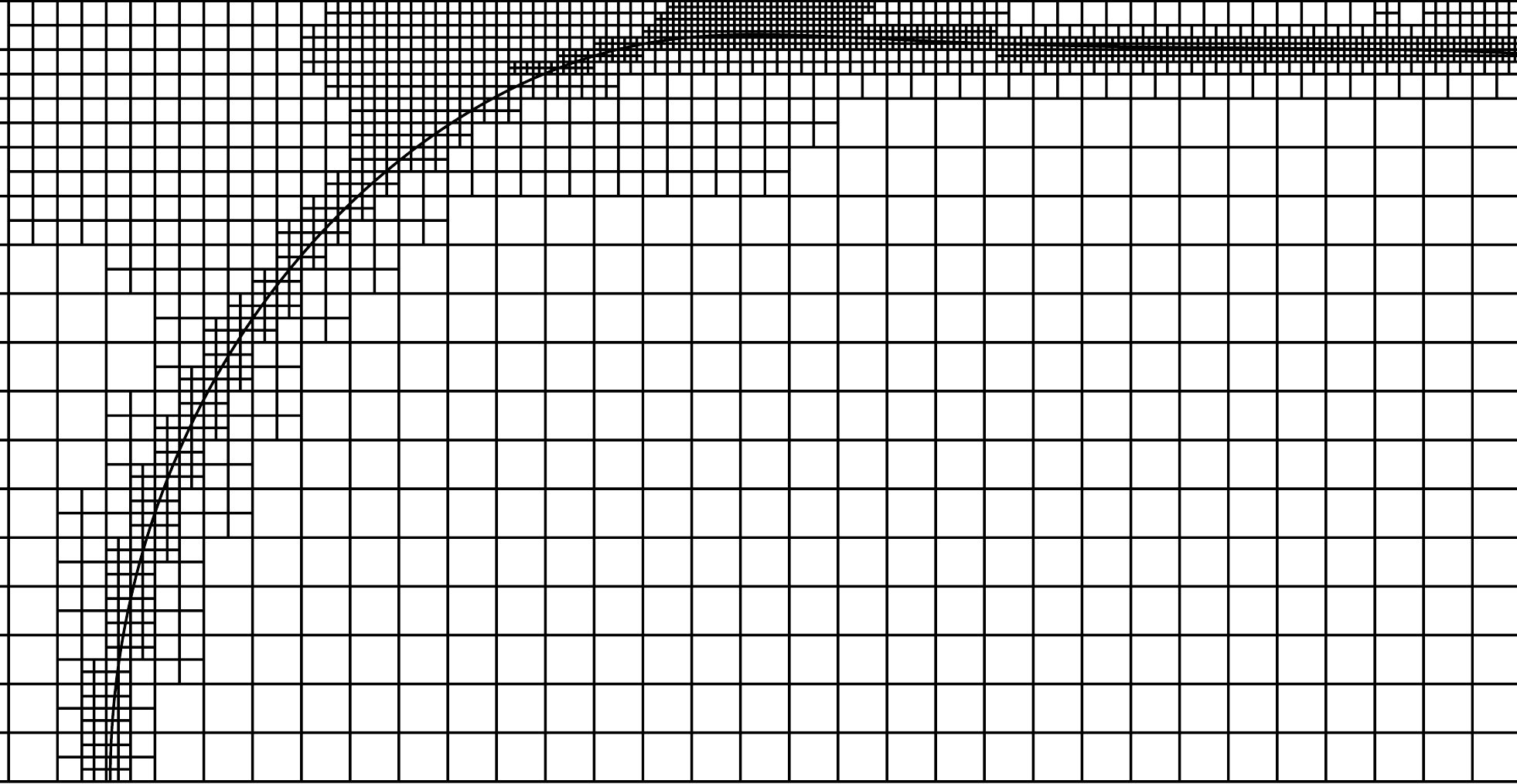}
\caption{(left) Typical bubble in the stationary state. (right) Grid refinement at the rear of the bubble.}
\label{fig2}
\end{figure}

\begin{eqnarray}
  \label{eq:5}
 {\partial U \over \partial t} + U \nabla U = - \nabla p + {1 \over R_e}\nabla^2 U + { 1 \over R_e C_a} \kappa \delta {\bar n}
\end{eqnarray}

where all dynamical variables are dimensionless and we define the Reynolds and Capillary numbers as
\begin{eqnarray}
R_e =  { \rho U H \over \mu}, \\
C_a = { \mu U_f \over \sigma}.
\end{eqnarray}
Over this communication we do not discriminate densities and viscosities for liquid and bubble as long as 
in the subsequent computations we fix the density and the viscosity ratios between the fluid and the bubble to one, then $\rho_f/\rho_b = 1$ and $\mu_f / \mu_b = 1$. Bretherthon \cite{Bretherton}  studied theoretically and experimentally the dynamic of a  bubble on a cylindrical  configuration. In the limit of small capillarity number based on the bubble velocity, it is shown that the ratio between the gap of the thin film of lubrication $h$ (between the wall and the bubble) and the typical height $H$of the channel  as well as the ratio between the bubble and mean fluid velocity scale both as $C_a^{2/3}$ :
\begin{eqnarray}
{ h \over H } \sim { U_b \over U_f} \sim C_a^{2/3}
\end{eqnarray}
the proportionality constant depends,  at least, on the geometrical configuration (planar, squared or cylindrical) and on the viscosity ratio. By inspecting the $h/H$ relation and its dependency into the capillary number $C_a$, it appears that the grid refinement plays an important role if we want to be able to capture the dynamics of the thin film. The Bretherton theory stands that thin film is very important as long as is the key point determining the bubble shape. The Figure \ref{fig2} (left) present the computed shape for a capillary number of $0.01$, which is in fully agreement with the theory. The Figure \ref{fig2} (right) shows the detail of the grid refinement at the rear of the bubble where the film is thinner. We can also note the specific grid refinement along the interface. 

The bubble velocity is evaluated by computing the $x$ position of the center mass of the bubble along the channel, $ { \int_S x \ c(x,t) \ dS \over \int_S dS}$ where $x$ is  the spatial position of each fraction $c$. 

\begin{figure}[htb]
\centering
\includegraphics[width=11cm]{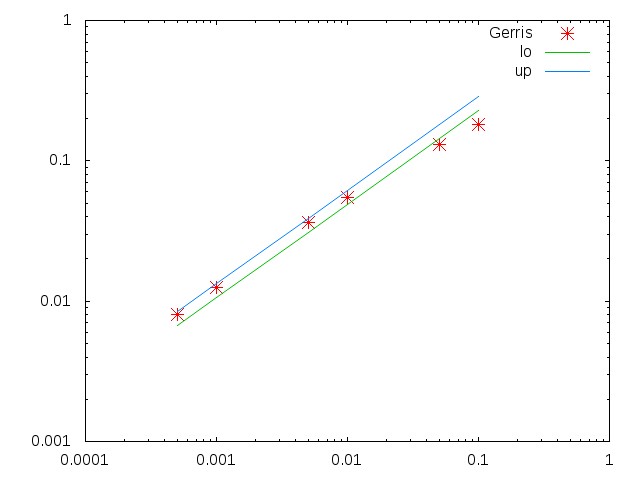}
\caption{Log-log scaling : ${U_b \over U_f } - 1$ as function of $C_a$. }
\label{fig3}
\end{figure}

We present now some quantitative and qualitative numerical results on a bubble flowing on a micro channel. For a bubble between parallel plates the analytical solutions are 
\begin{eqnarray}
{U_b \over U_f } \sim 1 + 0.643 \ (3 C_a)^{2/3} \ \ \ \mbox{or} \ \ \ \sim 1 + 0.51 \ (3 C_a)^{2/3} 
\end{eqnarray}
the second relation valid for very viscous drops (liquid-liquid interfaces). In our simulations we impose a Reynolds number of 0.1 and the only variable parameter is the capillarity number $C_a$. The Figure \ref{fig3} shows the log-log scaling of the excess of velocity ${U_b \over U_f } - 1$ as function of the capillarity number $C_a$ up to a capillarity number of $0.5 \ 10^{-4}$. The solid lines shows both limits from the later relations. These results are consistent with those of reference \cite{Afkhami2} where capillarity number are indeed greater. 

Experimental observations in a microfluidic setup \cite{ANR} show 
\begin{itemize}
\item that below a capillarity number of around $10^{-5}$ the film gap and the bubble velocity remain constant giving no dependency of the ratios $h/H$ and $U_b/U_f$ on the capillarity number, and
\item the bubble lost its Bretherton shape becoming more symmetric, like a pancake. 
\end{itemize}•
This phenomenon appears for $h$ of the order of tens of nanometers, an argument advanced as explanation is that in this region, near of the wall, the Van der Waals forces are not negligible pushing away the bubble by the apparition of an equilibrium film of constant width. The main difficult for including the Wan der Waals forces in this continuum approach is that's necessary resolving numerically a lot of scales, from the small, few nanometers for the equilibrium film, to the large ones, the channel width $H$. Including Van der Waals forces in a continuum approach was recently done using Gerris to impose the macroscopic contact angles from microscopic physics \cite{Afkhami}, the numerical computations were done locally and the wide scale range  problem was not matter of fact. Starting from the Lennard-Jones potential of two particles and doing some approximations we can add to the r.s.h. of the Navier-Stokes momentum equation (equation (\ref{eq:5})) the force $F(d)$ per unit of volume which depends only on the distance $d$ between the bubble interface and the wall 
\begin{eqnarray}
F(d) = { K \over d^*} \left[ m \left( {h^* \over d} \right)^{m+1} - n \left( {h^* \over d} \right)^{n+1} \right]
\label{vdw}
\end{eqnarray}•
where $K$ is a constant and $h^*$ is the equilibrium film thickness, $m=3$ and $n=2$. (details in reference \cite{Afkhami}). 

\begin{figure}[htb]
\centering
\includegraphics[width=0.7\textwidth]{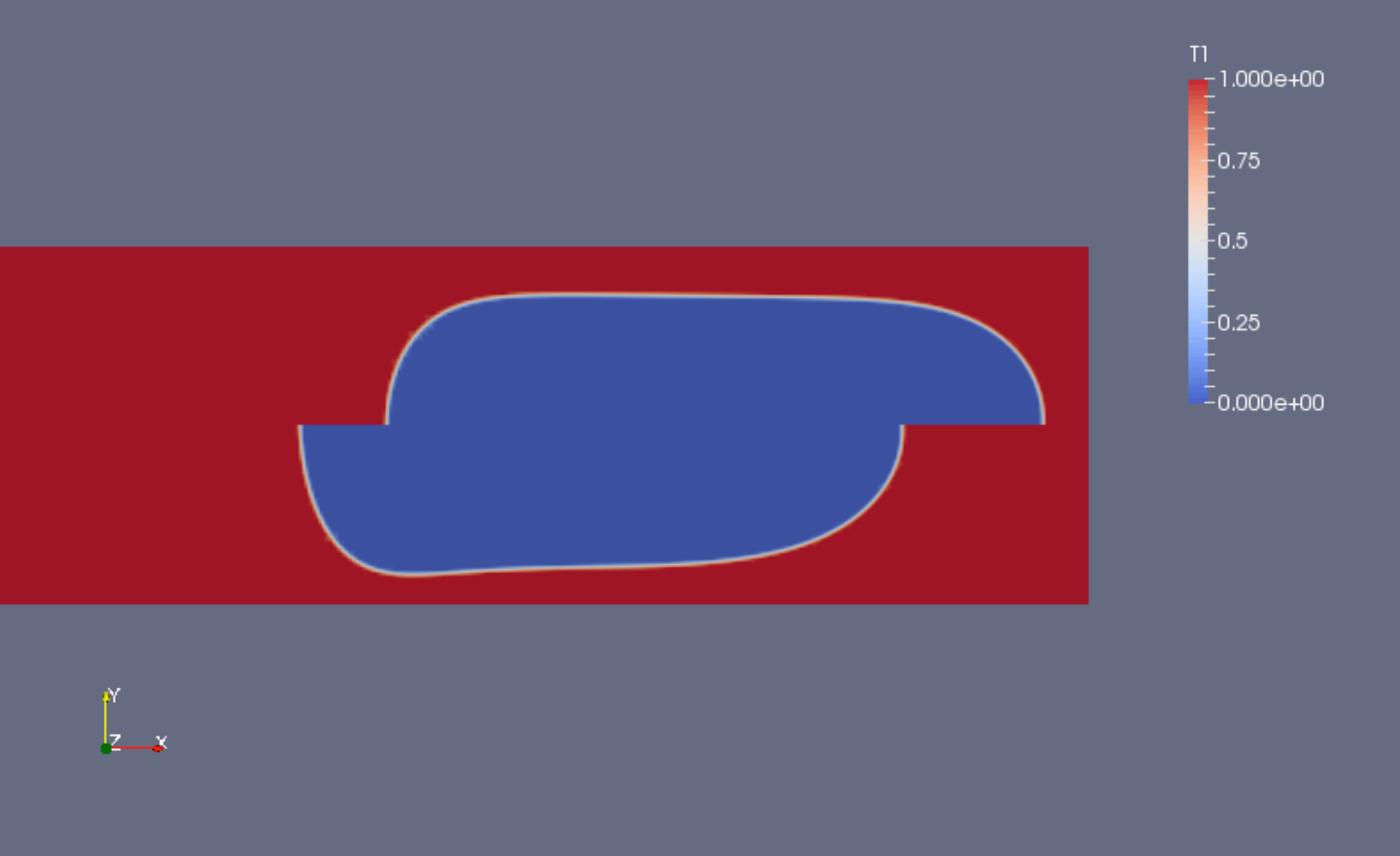}
\caption{Steady bubble shape and final position including  (upped side) or not (lower side) the Van der Waals forces.}
\label{fig4}
\end{figure}

To compare qualitatively this approach without resolving all the spatial scales we impose a large value of $h^*$ which is indeed not physical but the mechanism is still the same, pushing the bubble away from the wall. The Figure \ref{fig4} present two numerical simulations of a bubble flowing into a channel for $C_a = 0.01$ and $Re=0.1$ for the same final time  with (upped side) and with out (lower side) Van der Waals forces from equation (\ref{vdw}). We observe that (i) imposing the Van der Waals forces we found a larger gap $h$ and consequently a faster bubble velocity $U_b$, (ii) the bubble shape becomes a pancake like. 

\section{CONCLUSIONS}

We have presented numerical simulations of a bubble into a channel, the well behavior of the numerical implementation of the Navier-Stokes equations with a surface tension model was demonstrated by a comparison with Bretherton theory for very small capillary numbers were the scaling law in $C_a^{2/3}$ was validated. The quality of the numerical results are, in particular, a consequence of the grid refinement approach which allows computing the very thin films of liquid between the bubble and the wall. 
We have also implemented a Van der Waals like force and the imposed minimum gap gives numerical prediction in according with experimental observations.

\end{document}